\documentclass[useAMS,usenatbib,usegraphicx]{mn2e}
\title[Gravitational waves from N-body systems]
{
Can N-body systems generate periodic gravitational waves? 
}
\author[
T. Chiba, T. Imai and H. Asada]
{T. Chiba, T. Imai and H. Asada
\\ 
Faculty of Science and Technology, 
Hirosaki University, Hirosaki 036-8561, Japan}
\begin{document}

\date{Accepted  Received }

\pagerange{\pageref{firstpage}--\pageref{lastpage}} 

\pubyear{2006}

\maketitle

\label{firstpage}

\begin{abstract}
None of N-body gravitating systems have been considered to emit 
periodic gravitational waves because of their chaotic orbits 
when $N=3$ (or more). 
We employ a figure-eight orbit as a specific model for a 3-body system 
in order to illustrate that some of triple stars are capable of 
generating periodic waves. 
This illustration would imply that a certain class of N-body 
gravitating systems may be relevant to 
the gravitational waves generation. 
We show also that the total angular momentum of this 3-body 
system is not carried away by gravitational waves. 
A waveform generated by this system is volcano-shaped and 
thus different from that of a binary system. 
Finally, by evaluating the radiation reaction time scale, 
we give an order-of-magnitude estimate of merging event rates. 
The estimate 
suggests that figure-eight sources, which require carefully 
prepared initial states, may be too rare to detect. 
\end{abstract}


\section{Introduction}
Gravitational waves represent one of the great challenges 
of fundamental physics 
and will open a new window in the astronomy 
%\cite{MTW,Centrella,Mioa,Miob}. 
(e.g. Misner et al. 1973, Centrella 2003, Mio 2006a, 2006b).  
No one has ever detected directly the ripples 
in a curved space-time generated by accelerated masses. 
A (quasi-)periodic source is the most promising candidate 
for the first detection. 
Hence, (quasi-)periodic sources have been studied for the last few 
decades in numerous articles, where they assume 
a single star in rotation/oscillation 
%\cite{Kokkotas,Stergioulas} 
(e.g. Kokkotas and Schmidt 1999, Stergioulas 2003)
or double stars in binary motion 
%\cite{AF,Blanchet}. 
(e.g. Asada and Futamase 1997, Blanchet 2006). 
In particular, event rates of an inspiraling and finally merging 
compact binary 
%\cite{Curran,Heuvel,Arzoumanian,Kalogera,Kim,Burgay} 
(Curran et al. 1995, van den Heuvel et al. 1996, 
Arzoumanian et al. 1999, Kalogera et al. 2001, Kim et al. 2003, 
Burgay et al. 2003) 
are discussed by taking account of large-scale detectors 
such as LIGO, VIRGO, GEO600, TAMA300 and LISA 
%\cite{Centrella,Mioa,Miob}. 
(Centrella 2003, Mio 2006a, 2006b). 
Furthermore, gravitational waves generated by a binary system 
have been intensively studied by perturbations 
such as the post-Newtonian approximation 
%\cite{AF,Blanchet,FJS,BIW}, 
(e.g. Asada and Futamase 1997, Blanchet 2006, Faye et al. 2004, 
Berti et al. 2006), 
or by full numerical simulations 
for neutron stars 
%\cite{Shibata} 
(Shibata 1999) 
and for black holes 
%\cite{Pretorius,BBH06a,BBH06b,BBH06c}.  
(Pretorius 2005, Campanelli et al. 2006b, 
Baker et al. 2006, Diener et al. 2006).  
However, much less attention has been paid to N-body gravitating
systems because of their chaotic orbits when $N\geq 3$. 
%Here we report the discovery of Einstein's three-star legacy. 
There are existing works which are limited 
in the sense that they study a binary system affected 
by the existence of a third body, namely perturbations 
induced by the third object 
%\cite{ICTN,Wardell,CDHL}. 
(Ioka et al. 1998, Wardell 2002, Campanelli et al. 2006a). 
Here, one may ask whether or not 
N-body systems can generate (quasi-)periodic gravitational waves. 
The purpose of this paper is to answer this question. 

The main results of this paper are two; 
(1) For the first time as far as we know, 
we show that some of triple stars are capable of generating 
periodic waves if a certain condition is satisfied. 
This illustration would imply that a class of N-body 
gravitating systems may be relevant to 
the gravitational waves generation. 
(2) We show also that a solution employed as a specific model 
does not radiate the angular momentum.
A well-known example in which no angular momentum is radiated away 
is two non-spinning black holes in a head-on collision 
%\cite{Smarr,PP}, 
(Smarr 1979, Price and Pullin 1994), 
where this system is axisymmetric and thus has a rotational Killing
vector. Therefore, we can easily understand no angular momentum loss. 
On the other hand, the specific system employed in this paper 
has no Killing vector as seen below. 

Our paper is organized as follows. 
In this paper, we restrict ourselves within 
3-body systems for convenience. 
%First
In section 2, we assume the Newtonian gravity for motion of 
a 3-body system. 
In section 3, the waveform and the loss rate of the energy and 
the angular momentum are estimated by using 
the so-called quadrupole formula for simplicity. 
Finally, we discuss the the time scale for radiation reaction 
in order to estimate merging event rates.  

\section{Periodic Solutions for a three-body system} 
It is impossible to describe all the solutions 
to the 3-body problem even for the $1/r$ potential. 
The simplest periodic solutions for this problem 
was discovered by Euler (1765) and by Lagrange (1772). 
The Euler's solution is a collinear solution, 
in which the masses are collinear at every instant 
with the same ratios of their distances. 
The Lagrange's one is an equilateral triangle solution 
in which each mass moves in an ellipse in such a way 
that the triangle formed by the three bodies revolves.  
Built out of Keplerian ellipses, they are the only explicit
solutions. 
In these solutions, each mass moves on an ellipse 
(a circle for the equal mass case). 
Therefore, the associated gravitational waveform is 
a superposition of a waveform for each mass orbiting in an ellipse.  

An interesting solution that three bodies move 
periodically in a figure-eight was found firstly by Moore 
by numerical computations 
%\cite{Moore}. 
(Moore 1993). 
The existence of such a figure-eight orbit was proven 
by mathematicians Chenciner and Montgomery 
%\cite{CM}. 
(Chenciner and Montgomery 2000). 
This solution is shown to be stable in the Newtonian gravity 
%\cite{Simo,GMF}. 
(Simo 2002, Galan et al. 2002). 
The figure-eight seems unique up to scaling 
and rotation according to all numerical investigations, 
and at the end its unicity has been recently proven 
%\cite{Montgomery05}. 
(Montgomery 2005). 
Many efforts have been paid to reveal some properties 
such as the convexity in the figure-eight solution 
%\cite{KZ,FM}. 
(Kapela and Zgliczynski 2003, Fujiwara and Montgomery 2005). 
Furthermore, it is shown numerically that fourth, 
sixth or eighth order polynomial cannot express 
the figure-eight solution 
%\cite{Simo}. 
(Simo 2002). 
Nevertheless, no analytic expressions in closed forms for 
the figure-eight trajectory have been found up to now.  
Therefore, for the purpose of investigating gravitational waves, 
we numerically prepare the figure-eight orbit. 

For simplicity, we assume a 3-body system with each mass 
equal to $m$. 
Without loss of the generality, the orbital plane 
is taken as the $x-y$ plane. 
The position of each mass is denoted by 
$(x_A, y_A)$ for $A=1, 2, 3$. 
Figure 1 shows the figure-eight orbit, 
where we take the initial condition 
%as $(x_1, y_1)=(-x_2, -y_2)=(0.97000436, -0.24308753)$ 
as $(x_1, y_1)=(-x_2, -y_2)=(0.970, -0.243)$ 
and $(\dot{x}_3, \dot{y}_3)=(-2 \dot{x}_1, -2 \dot{y}_1)
=(-2 \dot{x}_2, -2 \dot{y}_2)
%=(-0.93240737, -0.86473146)$ , 
=(-0.932, -0.865)$ , 
where a dot denote the time derivative 
%\cite{Simo}. 
(Simo 2002). 
When one mass arrives at the knot (centre) of the figure-eight, 
$\ell$ is defined as  
a half of the separation between the remaining two masses. 
It is convenient to use $\ell$ instead of a distance between 
the knot and the apoapsis, because the inertial moment is 
expressed simply as $2 m \ell^2$. 
Clearly this system has no Killing vector as seen in Fig. 1. 
\begin{figure}
\includegraphics[width=8cm]{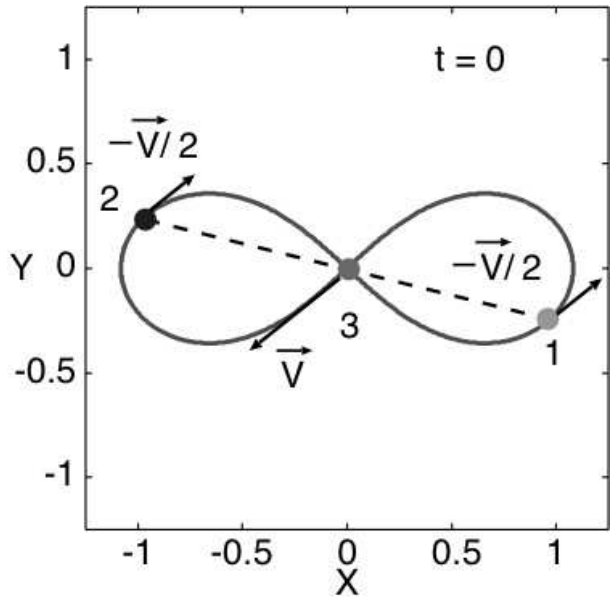}
\caption{
Three masses in a figure-eight at the initial time. 
Each mass is labeled by 1, 2 and 3. 
The initial velocity of each mass is denoted by an arrow. 
The distance between 1 and 3 is denoted by $\ell$. 
In this plot, we take $\ell=1$.
}
\end{figure}

The orbital period is estimated as 
$
%\begin{eqnarray}
%T&=&6.32591398 \sqrt{\frac{\ell^3}{Gm}} 
%T=6.32591 (Gm)^{-1/2} \ell^{3/2} 
T=6.33 (Gm)^{-1/2} \ell^{3/2} 
%\nonumber\\
%&\approx&
\approx
%10^4 
%\left(\frac{M_{\odot}}{m}\right)^{1/2} 
%\left(\frac{\ell}{R_{\odot}}\right)^{3/2} 
10^4 
(M_{\odot}/m)^{1/2} 
(\ell/R_{\odot})^{3/2} 
\: 
\mbox{sec.} , 
%\label{T}
%\end{eqnarray}
$
where $G$ denotes the Newtonian gravitational constant, 
and $M_{\odot}$ and $R_{\odot}$ are 
the solar mass and radius, respectively. 
Figure 2 shows a configuration of the three masses 
after a half of the period. 
The mass labeled by 3 starts at the knot and 
sweeps all the L.H.S. of the figure-eight 
during the first half of the period. 
\begin{figure}
\includegraphics[width=8cm]{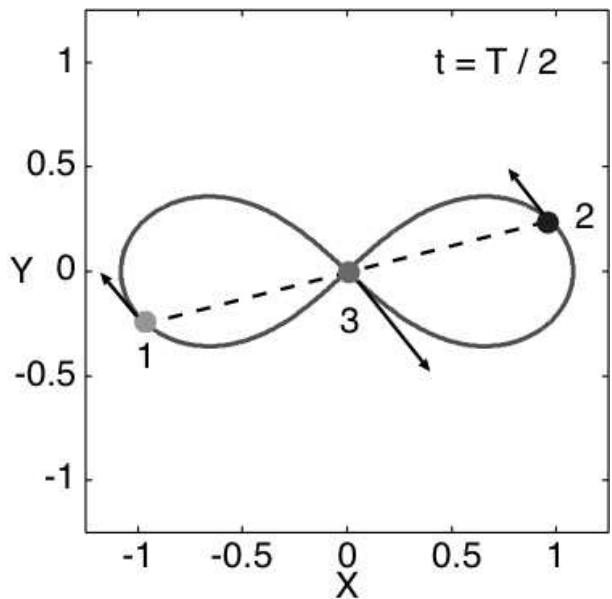}
\caption{
A figure-eight at $t=T/2$. 
The velocity of each mass is denoted by an arrow.
}
\end{figure}

\section{Gravitational Waves} 
The quadrupole moment $I_{ij}$ for $N$ masses is expressed as 
$
%\begin{equation}
I_{ij}=\sum_{A=1}^N m_A x_A^i x_A^j , 
%\label{Iij}
%\end{equation}
$
where $m_A$ denotes the $A$-th mass at the location $x_A^i$, and 
$i$ and $j$ run from 1 to 3 $(x^1=x, x^2=y, x^3=z)$. 
The reduced quadrupole %${I\!\!\!\!-}_{ij}$ 
$Q_{ij}$ 
is defined as 
$
%\begin{equation}
%{I\!\!\!\!-}_{ij}
Q_{ij}=I_{ij} - \delta_{ij} I_{kk}/3 .   
%\label{Ibar} 
%\end{equation}
$
In a wave zone, the gravitational waves denoted by $h^{TT}_{ij}$ 
become asymptotically 
% $
\begin{equation}
h^{TT}_{ij}=\frac{2G\ddot{Q}_{ij}}{rc^4}
+O\left(\frac{1}{r^2}\right) ,
% h^{TT}_{ij}=2G\ddot{\,I\!\!\!\!-}_{ij}/rc^4 
% +O(1/r^2) ,
%\label{hTT}
\end{equation}
% $
where $r$ is a distance from a source 
%\cite{MTW,Blanchet}. 
(e.g. Misner et al. 1973, Blanchet 2006). 
Here, $TT$ means the transverse ($h^{TT}_{ij} n^i=0$) 
and traceless ($h^{TT}_{ii}=0$), where $n^i$ denotes 
the unit vector of the direction of propagating gravitational waves. 
Figure 3 shows a volcano-shaped waveform from the figure-eight solution.

\begin{figure}
\includegraphics[width=8cm]{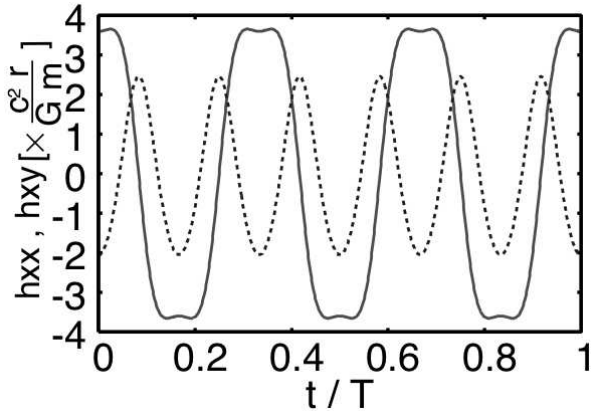}
\caption{
Gravitational waves from three masses in a figure-eight orbit. 
The dashed curve %(blue) 
and the solid one %(red) 
denote $h_{xx}^{TT}$ and $h_{xy}^{TT}$, respectively.
This plot is in contrast to that of an equal-mass binary, 
for which gravitational waves are expressed as a simple sine curve 
with one maximum and one minimum. 
On the other hand, there is a small valley between the nearest maxima 
and a small ridge between the nearest minima in Figure 3. 
At the bottom of the valley and the top of the ridge, 
one of the three masses passes the knot of the figure-eight, 
corresponding to Figs. 1 and 2, respectively. 
}
\end{figure}

The loss rate of the energy $(E)$ and the angular momentum $(L^i)$ 
is given by the quadrupole formula 
(Misner et al. 1973, Blanchet 2006). 
%as 
%\cite{MTW,Blanchet}. 
%\begin{eqnarray}
%\frac{dE}{dt}&=&-\frac{G}{5c^5} 
%<\dddot{\,I\!\!\!\!-}_{ij} \dddot{\,I\!\!\!\!-}_{ij}> , 
%\label{dE/dt}
%\\
%\frac{dL^i}{dt}&=&-\frac{2G}{5c^5} \epsilon^{ijk} 
%<\ddot{\,I\!\!\!\!-}_{ja} \dddot{\,I\!\!\!\!-}_{ak}> .  
%\label{dL/dt}
%\end{eqnarray}
%Here, the square bracket denotes the time averaging 
%and $\epsilon^{ijk}$ denotes the Levi-Civita symbol 
%defined as the completely antisymmetric unit tensor 
%as $\epsilon^{123}=1$. 
Because of $z_A^3=0$ for each mass, 
$dL^x/dt$ and $dL^y/dt$ vanish. 
%%%
In a case of the figure-eight solution, we can show numerically 
that $dL^z/dt$ vanishes. 
One may ask why the emitted waves carry away no total angular momentum, 
though each body is likely to lose its orbital angular momentum 
by gravitational waves. 
This apparent paradox happens because the total angular momentum 
of this 3-body system vanishes, while the angular momentum of 
each body does not necessarily, as seen from Fig. 1. 
As a result, gravitational waves carry away the angular momentum 
of each body, while it is impossible for this system 
to lose the total angular momentum as the sum. 
For instance in Fig. 1, at the initial time, 
the particle labeled by 3 has no orbital angular momentum. 
The particle 1 has the orbital angular momentum, 
which is carried away by gravitational waves. 
It is true of the particle 2 but in the opposite sign. 
Therefore, these changes in the orbital angular momentum cancel 
each other. 
This cancellation is expressed as 
$\sum_{A=1}^3 \dot{L}_A=0$, where $L_A$ denotes 
the angular momentum of each body. 
No angular momentum radiation as a consequence of such a cancellation 
is in contrast to the case of the head-on collision of two non-spinning 
black holes 
%\cite{Smarr,PP}. 
(Smarr 1979, Price and Pullin 1994). 
In the head-on collision case, each black hole moves 
without any orbital angular momentum with respect to 
the common centre of mass. In fact, the angular momentum 
of each black hole is not carried away 
($\dot{L}_B=0$ for each black hole). 
Clearly no total angular momentum is carried away. 

The energy loss rate is estimated as 
%$
\begin{equation}
\frac{dE}{dt}=1.2\times 10^{19} \left(\frac{m}{M_{\odot}}\right)^{5} 
%1.2\times 10^{19} (m/M_{\odot})^{5} 
\left(\frac{R_{\odot}}{\ell}\right)^{5} 
%(R_{\odot}/\ell)^{5} 
\: 
\mbox{erg/s} . 
%\label{dE/dt2}
\end{equation}
%$
As a result, the radiation reaction time scale is evaluated as 
%$
\begin{eqnarray}
t_{GW}&\equiv&\frac{E}{dE/dt} 
%t_{GW}\equiv E (dE/dt)^{-1} 
\nonumber\\
&=&0.13 \left(\frac{M_{\odot}}{m}\right)^{3} 
%=0.13 ({M_{\odot}}/{m})^{3} 
\left(\frac{\ell}{R_{\odot}}\right)^{4} 
%({\ell}/{R_{\odot}})^{4} 
\: 
\mbox{Gyr} ,  
\label{tGW}
\end{eqnarray}
%$
which gives a rough estimate of the collision time. 

It is worthwhile to mention that the dependence of 
$T$, $dE/dt$ and $t_{GW}$ on $m$ and $\ell$ 
is the same as that on the mass and separation of a binary, 
as implied also by a dimensional analysis. 

Let us consider a `compact' system of three black holes 
just before merging. 
To do so, we take $\ell$ as a much shorter distance, say 
a dozen of the Schwarzschild radius of each mass. 
Then, the orbital period of this system is 
of the order of a few milli-seconds.  
The frequency of the associated gravitational waves becomes a few kHz, 
around which the large-scale interferometric detectors 
(LIGO, VIRGO, GEO600, TAMA300) is most sensitive. 
We obtain the amplitude of the gravitational waves as 
%$
\begin{equation}
h^{TT}_{ij}\sim 10^{-17} 
\left(\frac{m}{10M_{\odot}}\right)
\left(\frac{10 M_{sch}}{\ell}\right)
\left(\frac{10\mbox{kpc}}{r}\right) ,  
%(m/10M_{\odot})
%(10 M_{sch}/\ell)
%(10\mbox{kpc}/r) ,  
%\label{amplitude}
\end{equation}
%$
where $M_{sch}$ denotes the Schwarzschild radius. 

Clearly we have to take account of relativistic corrections 
such as the back-reaction on the evolution of the orbit. 
If the system is secularly stable against the gravitational radiation, 
one might see probably a shrinking ($\dot\ell < 0$) figure-eight orbit 
as a consequence of decreasing the total energy 
($\dot E<0$)and keeping the vanishing total angular momentum ($L=0$). 
Indeed, a chirp signal caused by the back-reaction 
at the inspiraling phase plays a crucial role 
in the gravitational waves detection. 
Furthermore, the numerical relativity seems necessary for evolving 
the system in its merging phase after the last stable orbit.

\section{Concluding Remarks} 
A figure-eight orbit for a 3-body system 
illustrated that some of triple stars are capable of generating 
periodic waves. 
We showed also that the total angular momentum of this 3-body 
system is not carried away by gravitational waves. 
A waveform generated by this system is volcano-shaped and 
thus different from that of a binary system. 
Our result will call our further attention to, 
for instance, %the backreaction on the evolution of the orbit and 
N=4 (or more) cases. 

Before closing this paper, we mention the possibility 
of gravitational waves from 3-body systems in a `periodic' orbit 
such as a figure-eight. 
Clearly the possibility seems extremely low though it is quite uncertain. 
A quick order-of-magnitude estimate is made as follows.  
As a new outcome of binary-binary scattering, the figure-eight
orbit was discussed for presenting a way of detecting such an orbit 
in numerical computations 
%\cite{Heggie}. 
(Heggie 2000). 
According to the numerical result, the probability of 
the formation of figure-eight orbits is a tiny fraction of one
percent. 
Here, we assume rather optimistically 
one per galaxy, which implies $10^6$ figure-eight systems 
within 100Mpc. 
If $\ell$ distributes randomly with its mean of O(1AU), 
we use Eq. $(\ref{tGW})$ to obtain $t_{GW} \sim 10^8$ Gyr 
for $\ell \sim 1$AU. 
As a result, merging event rates are of the order of $10^{-2}$ 
event/Gyr within 100 Mpc, which is extremely low.  
Therefore, the above estimate suggests that 
figure-eight sources may be less relevant 
to LIGO and other detectors.

\section*{Acknowledgments}
We would like to thank Professor M. Shibata 
for useful comments.
We would like to thank Professor S. Kuramata and 
Professor M. Kasai for continuous encouragement.

\bsp

\label{lastpage}


\begin{thebibliography}{99}
\bibitem{Arzoumanian}
Arzoumanian Z., Cordes J. M., Wasserman I., 
%Pulsar Spin Evolution, Kinematics, and the Birthrate of Neutron Star
%Binaries. 
1999, ApJ., 520, 696 
\bibitem{AF}
Asada H., Futamase T., 
%post-Newtonian Approximation 
%--- Its foundation and applications. 
1997, Prog. Theor. Phys. Suppl., 128, 123 
\bibitem{BBH06b}
Baker J. G., Centrella J., Choi D. I., Koppitz M., vanMeter J.,  
%Gravitational-wave extraction from an inspiraling 
%configuration of merging black holes. 
2006, Phys. Rev. Lett., 96, 111102 
\bibitem{BIW}
Berti E., Iyer S., Will C. M., 
2006, Phys. Rev. D, 74, 061503(R) 
\bibitem{Blanchet}
Blanchet L., 
%Gravitational radiation from post-Newtonian sources and 
%inspiralling compact binaries. 
2006, Living Rev., No. 4 
\bibitem{Burgay}
Burgay M. et al.,  
%An increased estimate of the merger rate of double neutron stars from 
%observations of a highly relativistic system. 
2003, Nature, 426, 531 
\bibitem{CDHL}
Campanelli M., Dettwyler M., Hannam M., Lousto C. O., 
2006a, Phys. Rev. D, 74, 087503 
\bibitem{BBH06a}
Campanelli M., Lousto C. O., Marronetti P., Zlochower Y., 
%Accurate evolutions of orbiting black-hole binaries 
%without excision. 
2006b, Phys. Rev. Lett., 96, 111101 
\bibitem{Centrella}
Centrella J. (ed.), 2003, 
{\it The astrophysics of gravitational wave sources}, 
AIP conference proceedings, 686 
\bibitem{CM} 
Chenciner A., Montgomery R., 
%A remarkable periodic solution of the three-body problem 
%in the case of equal masses. 
2000, Ann. Math., 152, 881 
\bibitem{Curran}
Curran S. J., Lorimer D. R., 
%Pulsar Statistics. Part 3: Neutron Star Binaries. 
1995, MNRAS, 276, 347 
\bibitem{BBH06c}
Diener P. et al. 
%Accurate evolution of orbiting binary black holes. 
2006, Phys. Rev. Lett., 96, 121101 
\bibitem{FJS}
Faye G., Jaranowski P., Schafer G., 
2004, Phys. Rev. D, 69, 124029 
\bibitem{FM}
Fujiwara T., Montgomery R., 
%Convexity in the figure eight solution to the three-body problem. 
2005, Pacific J. Math., 219, 271 
\bibitem{GMF}
Galan J., Munoz-Almaraz F. J., Freire E., 
Doedel E., Vanderbauwhede A., 
2002, Phys. Rev. Lett., 88, 241101 
\bibitem{Heggie}
Heggie D. C., 
2000, MNRAS, 318, L61 
\bibitem{ICTN}
Ioka K., Chiba T., Tanaka T., Nakamura T., 
1998, Phys. Rev. D, 58, 063003 
\bibitem{Kalogera}
Kalogera V., Narayan R., Spergel D. N., Taylor J. H., 
%The Coalescence Rate of Double Neutron Star Systems. 
2001, ApJ., 556, 340 
\bibitem{KZ}
Kapela T., Zgliczynski P., 
%The existence of simple choreographies for the N-body problem 
%--- a computer-assisted proof. 
2003, Nonlinearity, 16, 1899 
\bibitem{Kim}
Kim C., Kalogera V., Lorimer D. R., 
%The Probability Distribution of Binary Pulsar Coalescence Rates. I. 
%Double Neutron Star Systems in the Galactic Field. 
2003, ApJ., 584, 985 
\bibitem{Kokkotas}
Kokkotas K., Schmidt B., 
%Quasi-normal modes of stars and black holes. 
1999, Living Rev., No. 2
\bibitem{Mioa}
Mio N. (ed.), 2006a, 
{\it Sixth Edoardo Amardi Conference on Gravitational Waves}, 
Journal of Physics: Conference Series, 32 
\bibitem{Miob}
Mio N. (ed.),  2006b, 
{\it Selected papers from the Sixth Edoardo Amardi Conference 
on Gravitational Waves}, 
Class. Quant. Grav., (special issue) 23 
\bibitem{MTW}Misner C. W., Thorne K. S., Wheeler J. A., 1973, 
{\it Gravitation}, 
(Freeman, New York) 
\bibitem{Montgomery05} 
Montgomery R., 
%Fitting hyperbolic pants to a three-body problem.  
2005, Ergodic Theory and Dynamical Systems, 25, 921 
\bibitem{Moore} 
Moore C., 
%Braids in Classical Dynamics. 
1993, Phys. Rev. Lett., 70, 3675 
\bibitem{Pretorius}
Pretorius F., 
%Evolution of binary black-hole spacetimes. 
2005, Phys. Rev. Lett., 95, 121101 
\bibitem{PP}
Price R. H., Pullin J., 
1994, Phys. Rev. Lett., 72, 3297 
\bibitem{Shibata}
Shibata M., 
%Fully general relativistic simulation of coalescing binary 
%neutron stars: Preparatory tests. 
1999, Phys. Rev., 60, 104052 
\bibitem{Simo}
Simo C., 
%Dynamical properties of the figure eight solution 
%of the three-body. 
2002, Contemp. Math., 292, 209 
\bibitem{Smarr}
Smarr L. L., 1979, 
in {\it Sources of Gravitational Radiation}, 
(Cambridge University Press, Cambridge)
\bibitem{Stergioulas}
Stergioulas N., 
%Rotating stars in relativity. 
2003, Living Rev., No. 3 
%\bibitem{Sasaki}
%Sasaki, M. \& Tagoshi, H. 
%Analytic black hole perturbation approach to gravitational radiation. 
%{\it Living Rev.} No. {\bf 6}, (2003). 
\bibitem{Heuvel}
van den Heuvel E. P. J., Lorimer D. R., 
%On the galactic and cosmic merger rate of double neutron stars. 
1996, MNRAS, 283, L37 
\bibitem{Wardell}
Wardell Z. E., 
2002, MNRAS, 334, 149 
\end{thebibliography}
\end{document}